\documentclass[11pt,a4paper,twocolumn]{scrartcl}

\usepackage[utf8]{inputenc}
\usepackage{geometry}
\usepackage{glossaries}
\usepackage{ifthen}
\usepackage[detect-all,binary-units]{siunitx}
\usepackage{amsmath}
\usepackage{pgfplotstable}
\usepackage{xfrac}
\usepackage{float}
\usepackage[]{algorithmic}
\usepackage[font={small}]{caption}
\usepackage{authblk}
\usepackage{import}
\usepackage{pgfplotstable}
\usepackage{tikz}
\usepackage{hyperref}
\usepackage{soul}
\usepackage[notextcomp]{stix}
\usepackage{flushend}
\usepackage[]{todonotes}
 
\usetikzlibrary{external}
\newfloat{algorithm}{t}{lop}
\floatname{algorithm}{Algorithm}

\newcommand{\tb}[1]{\textbf{#1}}
\newcommand{\thetaw}{\theta_\text{w}}
\newcommand{\winit}{w_\text{init}}
\newcommand{\taustdp}{\tau_\text{STDP}}
\newcommand{\fmax}{f_\text{max}}

\hypersetup{colorlinks,%
	citecolor=blue,%
	filecolor=black,%
	linkcolor=black,%
	urlcolor=black
}

\usetikzlibrary{shapes.misc,shapes}
\usetikzlibrary{decorations.pathreplacing}
\usetikzlibrary{positioning,fit,calc}

\usetikzlibrary{arrows.meta,decorations.pathreplacing,fadings,shapes,arrows}

\definecolor{blue}{RGB}{0,91,130}
\definecolor{lightblue}{RGB}{110,159,189}
\definecolor{red}{RGB}{185,70,60}
\definecolor{lightred}{RGB}{198,141,132}
\definecolor{green}{RGB}{125,150,110}
\definecolor{lightgreen}{RGB}{164,181,153}
\definecolor{orange}{HTML}{D7AA50}
\definecolor{purple}{HTML}{7A68A6}

\def\FigureScale{1.0}

\makeatletter
\def\relativepath{\import@path}
\makeatother

\setcapindent{0pt}

\deffootnote[0.3em]{0.0em}{0em}{\textsuperscript{\thefootnotemark}}

\makeatletter
\patchcmd{\@maketitle}{\LARGE \@title}{\fontsize{16}{19.2}\selectfont\@title}{}{}
\makeatother

\setkomafont{title}{\normalfont\Huge}
\setkomafont{section}{\normalfont\Large}
\setkomafont{subsection}{\normalfont\large}

\newacronym{ppu}{PPU}{plasticity processing unit}
\newacronym{padi}{PADI}{parallel driver interface}
\newacronym{aer}{AER}{address event representation}

\newacronym{sram}{SRAM}{static random-access memory}
\newacronym{dram}{DRAM}{dynamic random-access memory}
\newacronym{adc}{ADC}{analog-to-digital converter}
\newacronym{cadc}{CADC}{column-parallel analog-to-digital converter}
\newacronym{dac}{DAC}{digital-to-analog converter}
\newacronym{dut}{DUT}{design under testing}
\newacronym{asic}{ASIC}{application-specific integrated circuit}
\newacronym{fpga}{FPGA}{field-programmable gate array}
\newacronym{vlsi}{VLSI}{very-large-scale integration}
\newacronym{simd}{SIMD}{single instruction, multiple data}
\newacronym{ocp}{OCP}{Open Core Protocol}
\newacronym{soc}{SoC}{system on a chip}
\newacronym{sta}{STA}{static timing analysis}
\newacronym{gals}{GALS}{globally asynchronous locally synchronous}
\newacronym{dpi}{DPI}{SystemVerilog direct programming interface}
\newacronym{pll}{PLL}{phase-locked loop}
\newacronym{fifo}{FIFO}{First-In-First-Out Buffer}

\newacronym{psc}{PSC}{post-synaptic current}
\newacronym{psp}{PSP}{post-synaptic potential}
\newacronym{stp}{STP}{short-term plasticity}
\newacronym{stdp}{STDP}{spike-timing-dependent plasticity}
\newacronym{rstdp}{R-STDP}{reward-modulated spike-timing-dependent plasticity}
\newacronym{lif}{LIF}{leaky integrate-and-fire}
\newacronym{adex}{AdEx}{adaptive exponential leaky integrate-and-fire}

\newacronym{mc}{MC}{Monte Carlo}

\usepackage[authoryear,round]{natbib}
\title{Structural plasticity on an accelerated analog neuromorphic hardware system}

\date{\today}

\begin{document}

\author[$\star$,1]{Sebastian Billaudelle}
\author[$\star$,1]{Benjamin Cramer}
\author[3,1]{\\Mihai A. Petrovici}
\author[1]{Korbinian Schreiber}
\author[2]{David Kappel}
\author[$\S$,1]{\\Johannes Schemmel}
\author[$\S$,1,$\dagger$]{Karlheinz Meier}
\affil[$\star$]{Authors with equal contribution}
\affil[$\S$]{Shared senior authorship}
\affil[$\dagger$]{Deceased}
\affil[1]{Kirchhoff-Institute for Physics, Heidelberg University}
\affil[2]{Georg-August-Universität Göttingen}
\affil[3]{Department of Physiology, University of Bern}

\maketitle

\begin{abstract}
\noindent\bfseries
In computational neuroscience, as well as in machine learning, neuromorphic devices promise an accelerated and scalable alternative to neural network simulations.
Their neural connectivity and synaptic capacity depends on their specific design choices, but is always intrinsically limited.
Here, we present a strategy to achieve structural plasticity that optimizes resource allocation under these constraints by constantly rewiring the pre- and postsynaptic partners while keeping the neuronal fan-in constant and the connectome sparse.
In particular, we implemented this algorithm on the analog neuromorphic system BrainScaleS-2.
It was executed on a custom embedded digital processor located on chip, accompanying the mixed-signal substrate of spiking neurons and synapse circuits.
We evaluated our implementation in a simple supervised learning scenario, showing its ability to optimize the network topology with respect to the nature of its training data, as well as its overall computational efficiency.

\end{abstract}
\vspace{0.5em}

{
\bfseries
\itshape
\noindent
structural, plasticity, receptive fields, BrainScaleS, mixed-signal, neuromorphic, spiking, neural networks

}

\glsresetall

\section{Introduction}
\noindent
Experimental data shows that plasticity in the brain is not limited to changing only the strength of connections.
The structure of the connectome is also continuously modified by removing and creating synapses \citep{grutzendler2002long,zuo2005development,bhatt2009dendritic,holtmaat2009experience,xu2009rapid}.
Structural plasticity allows the nervous system to reduce its spatial and energetic footprint by limiting the number of fully expressed synaptic spine heads and maintaining sparsity \citep{knoblauch2016structural}.
The lifetime of dendritic spines, involved at least in excitatory projections, varies dramatically \citep{trachtenberg2002}.

The process of spine removal depends on the spine head size: smaller spines are removed while larger ones persist \citep{holtmaat2005transient,holtmaat2006experience}.
At the same time, new spines are continuously created.
The spine volume also shows a strong correlation with the amplitude of the respective synaptic currents \citep{matsuzaki2001dendritic}, hence suggesting a coupling of a connection's lifetime and its synaptic efficacy.

Neuromorphic devices implement novel computing paradigms by taking inspiration from the nervous system.
With the prospect of solving shortcomings of existing architectures, they often also inherit some restrictions of their biological archetypes.
The exact form and impact of these limitations depend on the overall design and architecture of a system.
Ultimately however, all physical information processing systems, with neuromorphic ones making no exception, have to operate on finite resources.

For most neuromorphic systems, synaptic fan-in is -- to various degrees -- one of these limited resources.
This applies to analog as well as digital platforms, especially when they implement fast on-chip memory.
For example, TrueNorth and ODIN both allocate fixed memory regions for 256 synapses per neuron \citep{akopyan2015,frenkel20180}.
Loihi imposes an upper limit of 4096 individual presynaptic partners per group of up to 1024 neurons located on a single core \citep{davies2018}.
In contrast, the digital neuromorphic multi-core platform SpiNNaker \citep{furber2013} allows to trade the number of synapses per neuron against overall network size or simulation performance.
In general, digital systems often make use of time-multiplexed update logic, and hence can be designed to alleviate the issue of a limited fan-in by increasing memory size -- albeit at the cost of prolonged simulation times.

Analog and mixed-signal systems mostly do not allow this trade-off, because their synapses are implemented physically, and therefore often constitute a fixed resource.
Examples include DYNAP-SEL \citep[providing 64 static synapses per neuron on four of its cores and 256 learning synapse circuits on a fifth core,][]{moradi2018scalable,thakur2018large}, Spikey \citep[256 synapses per neuron,][]{schemmel2007}, and BrainScaleS-1 \citep[220 synapses per neuron,][]{schemmeliscas2010}.
For this manuscript, we have used a prototype system of the BrainScaleS-2 architecture with 32 synapses per neuron.
At full scale, a single \gls{asic} features 256 synapses per neuron, with the additional option of merging multiple neuron circuits to larger logical entities in order to increase their overall fan-in \citep{aamir2018adex}, similarly to its predecessor BrainScaleS-1 \citep{schemmeliscas2010}.

The above list of neuromorphic systems is certainly not exhaustive, but it hints towards an ubiquitous issue of limited synaptic resources.
A promising way to address this issue is to, once again, draw inspiration from the biological nervous system by supporting the emulation of sparse networks.
In the field of machine learning, \citet{denil2013predicting} found that multi-layer networks in fact express many redundant connections for a wide range of common machine learning datasets.
Consequently, new training schemes have been developed to incorporate the concept of sparsity during training~\citep{bellec2017deep,wen2016learning}, allowing to compress large, fully connected networks into smaller, sparse representations without a significant loss in performance.
In order to make neuromorphic devices amenable to such compressed models, it appears desirable to deeply anchor mechanisms for sparse networks and structural plasticity in the design of the systems themselves.

In this paper we present an efficient structural plasticity mechanism and an associated on-chip implementation thereof for the BrainScaleS-2 system, directly exploiting the synapse array's architecture.
It leverages the fact that the network connectivity is partially defined and resolved within each synapse, which is enabled by local event filtering.
This design choice promotes the mapping of sparse network graphs to the synapse matrix.
The update algorithm is implemented on the embedded plasticity processor, which directly interfaces the synaptic memory through a vector unit.
This near-memory design allows efficient parallel updates to the network's topology and weights.

Our approach enables fully local learning in a sparse connectome while inherently keeping the synaptic fan-in of a neuron constant.
We further demonstrate its ability to optimize the network topology by forming clustered receptive fields and study its robustness with respect to sparsity constraints and choice of hyperparameters.
While enabling an efficient, task-specific allocation of synaptic resources through learning, we also point out that our implementation of structural plasticity is computationally efficient in itself, requiring only a small overhead compared to the computation of, e.g., synaptic weight updates.

\begin{figure}
	\begin{center}
		\scalebox{\FigureScale}{
			\begin{tikzpicture}
				\node[anchor=north west] (a) at (0.0,0) {\tikzset{block/.style={font={\rmfamily\footnotesize},align=center}}%
\tikzset{box/.style={draw=black!90}}%
\tikzset{block label/.style={fill=white,font={\rmfamily\footnotesize},inner sep=0.05cm}}%
\begin{tikzpicture}[
	baseline=(anc.south),
            font={\rmfamily \small},
            scale=0.7,
            >=stealth,
            transform shape,
	    line width=1.0\pgflinewidth,
	    anchor=center
        ]
        \pgfdeclarelayer{background layer}
        \pgfsetlayers{background layer,main}

	\node[block,box,minimum width=3.0cm,minimum height=2.0cm] (syn) at (0,0) {};
	\node[block,minimum width=3.0cm,minimum height=0.5cm,below=0.2cm of syn] (nrn) {};
	\node[block,box,minimum width=3.0cm,minimum height=0.5cm,below=0.1cm of nrn] (cm) {parameter storage};

	\node[block,minimum width=0.5cm,minimum height=2.0cm,left=0.05cm of syn] (drv) {};
	\node[block,minimum width=0.2cm,minimum height=2.0cm,left=0.05cm of drv] (padi) {};
	
	\node[block,box,minimum width=3.0cm,minimum height=0.3cm,above=0.1cm of syn] (cadc) {CADC};
	\node[block,box,minimum width=3.0cm,minimum height=0.5cm,above=0.5cm of cadc] (ppu) {PPU};

	\node[anchor=north west,gray,rotate=90] (anc) at (cm.south east) {\scriptsize analog network core};
	\node[block,box,gray,dashed,fit={(padi) (cadc) (anc)},inner sep=0.1cm] (anc bound) {};

	\node[block,box,anchor=south east,minimum width=1.7cm,minimum height=0.8cm] (event)
		at ($(nrn.south west -| anc bound.west) + (-0.2,0.2)$) {Event \\ Handling};
	\node[block,box,anchor=south east,minimum width=2.2cm,minimum height=0.4cm] (link)
		at ($(anc bound.south west) - (0.2,0.0)$) {I/O};
	\node[block,box,anchor=north east,minimum width=1.7cm,minimum height=1.2cm] (mem)
		at ($(anc bound.north west) - (0.2,0.0)$) {Config.\\Memory\\Controllers};

	\foreach \x in {0,1,...,5} {
		\draw[] (nrn.south west) ++ (\x*0.5,0.0) ++ (0.25,0.25) circle (0.2cm);
		\draw[gray,stealth-] (syn.south west) ++ (\x*0.5+0.25,-0.22) -- ++(0.0,2.22);
		
		\foreach \y in {0,1,...,3}
			\draw[gray] (syn.south west) ++ (\x*0.5+0.25,0.25+\y*0.5) circle (0.02cm);
	}

	\foreach \y in {0,1,...,3} {
		\draw[] (drv.south west) ++ (0.0,\y*0.5) ++ (0.1,0.05) -- ++(0.0,0.4) -- ++(0.3,-0.2) -- cycle;
		\draw[gray] (syn.south west) ++ (-0.1,\y*0.5+0.25) -- ++(3.0,0.0) -- ++(0.1,0.0);
	}

	\draw ($(event.east |- drv.south west) + (0.0,0.25)$) -- ($(drv.south west) + (-0.2,0.25)$) -- ++(0.0,1.5);
	\draw[,stealth-] (event.east |- nrn.west) -- (nrn.west);
	\draw[,stealth-stealth] (link.north -| event) -- (event);
	\foreach \y in {0,1,...,3}
		\draw[stealth-,line cap=rect] (drv.south west) ++ (0.1,\y*0.5+0.25) -- ++(-0.3,0.0);
	\foreach \x in {0,1,...,5}
		\draw[stealth-stealth,line cap=rect] (ppu.south west) ++ (\x*0.5,0) ++ (0.25,0.0) -- ++(0.0,-0.35);

	\draw[stealth-stealth] (event.west) -- ++(-0.25,0.0) coordinate (x) -- (link.north -| x);
	\draw[-stealth] (x) -- (x |- ppu) -- (ppu);
	\draw[-stealth] (x) -- (x |- mem) -- (mem);
	\draw[-stealth] (mem.east) ++ (0.0, 0.1) -- ++(0.2,0.0);
	\draw[stealth-] (mem.east) ++ (0.0,-0.1) -- ++(0.2,0.0);
	
	\node[block label] at (syn) {synapse array};
	\node[block label] at (nrn) {neurons};
  \end{tikzpicture}};
				\node[anchor=north west] (b) at (5.5,0) {\includegraphics[width=2.9cm]{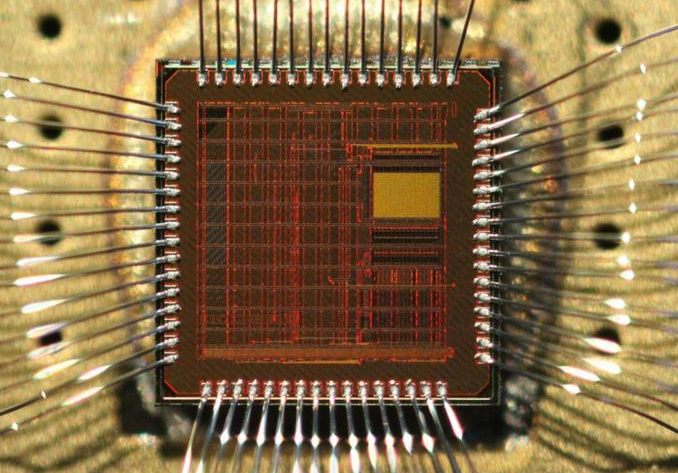}};
				\node[anchor=north west] (c) at (5.5,-2.25) {\includegraphics[width=2.9cm]{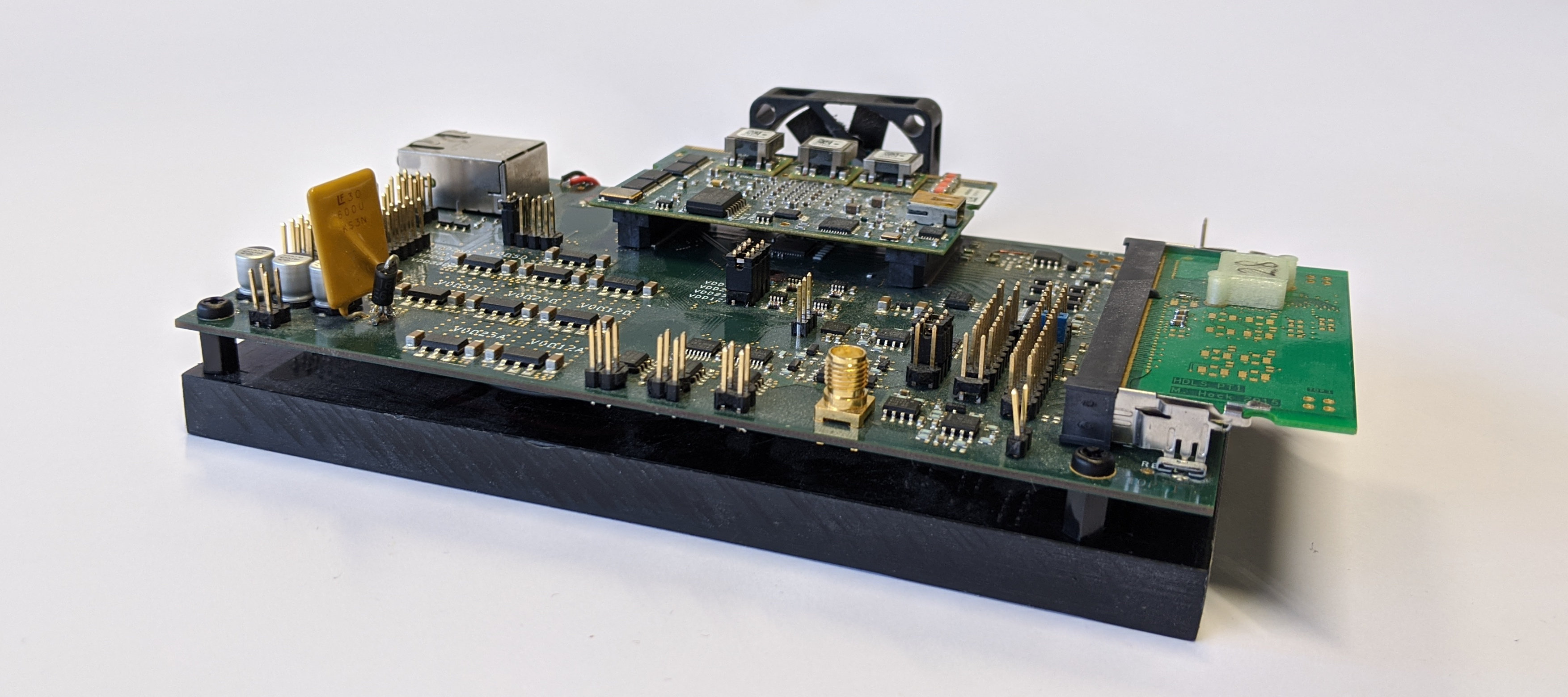}};

				\node at ($(a.north west) + ( 0.10,-0.25)$) {\small\bfseries A};
				\node at ($(b.north west) + (-0.04,-0.25)$) {\small\bfseries B};
				\node at ($(c.north west) + (-0.04,-0.25)$) {\small\bfseries C};

				%

			\end{tikzpicture}
		}
	\end{center}
	\caption{
	    \tb{BrainScaleS-2 prototype \acrshort{asic}.}
	    \tb{(A)}
        Block-level schematic.
	The analog neuromorphic core contains neuronal and synaptic circuits, which are accompanied by, inter alia, an analog parameter storage and the \acrshort{cadc} for digitizing synaptic correlation data.
	    It is surrounded by digital logic which interfaces the full-custom circuits and handles configuration data as well as spike traffic.
	    The \acrshort{ppu} is closely attached to the analog core, allowing it to access synaptic weights, address labels, and digitized correlation traces from the \acrshort{cadc}.
	   \tb{(B)} Photograph of the BrainScaleS-2 prototype \acrshort{asic}.
	   \tb{(C)} Experimental setup.
	    }
	\label{fig:arch}
\end{figure}
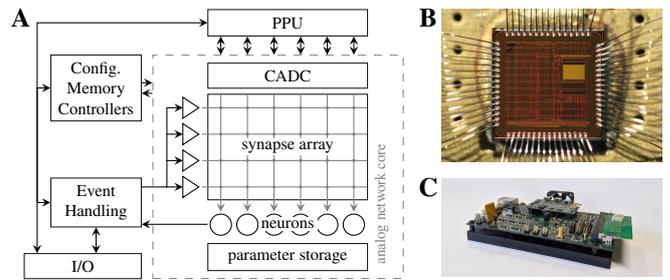

\section{Methods}
The BrainScaleS-2 architecture, which we discuss in section~\ref{ssec:arch}, provides all features required to implement flexible on-chip plasticity rules, including our proposed mechanism for structural reconfiguration.
Section~\ref{ssec:pruning} describes the algorithm for pruning and reassignment of synapses as well as an optimized implementation thereof.
This structural plasticity scheme can be coupled with various weight dynamics.
In this work, we employ a correlation-based weight update rule, which is described in section~\ref{ssec:updates}.
The combination of both is tested in a supervised classfication task, as outlined in section~\ref{ssec:task}.

\subsection{BrainScaleS-2 architecture}
\label{ssec:arch}
BrainScaleS-2 is a family of mixed-signal neuromorphic chips implemented in a \SI{65}{\nano\meter} process (Fig.~\ref{fig:arch}).
It is centered around an analog neural network core implementing neuron and synapse circuits that behave similarly to their biological archetypes.
State variables such as membrane potentials and synaptic currents are physically represented in the respective circuits and evolve in continuous time.
Leveraging the intrinsic capacitances and conductances of the technology, time constants of neuron and synapse dynamics are rendered 1000 times smaller compared to typical values found in biology.
This thousandfold acceleration facilitates the execution of time-consuming tasks, such as performing high-dimensional parameter sweeps, the investigation of learning and metalearning, or statistical computations requiring large volumes of data \citep{cramer2019control, bohnstingl2019neuromorphic}.

The analog core features 32 silicon neurons%
\footnote[1]{
    Later versions of the system feature 512 neuron circuits with adaptive-exponential \acrshort{lif} dynamics and inter-compartmental conductances.
    Each neuron is connected to 256 synapse circuits.
    Support for conductance-based synapses is planned for future versions.
    }
\citep{aamir2018lifarray} implementing \gls{lif} dynamics
$C_\text{m} \dot V_\text{m} = -g_\text{l}(V_\text{m} - E_\text{l}) + I_\text{syn},$ where $V_\text{m}$ represents the membrane potential, $C_\text{m}$ the membrane capacitance, $g_\text{l}$ the leak conductance, and $E_\text{l}$ the resting potential.
Synaptic currents $I_\text{syn}$ are modeled as superpositions of spike-triggered exponential kernels.
The membrane is connected to a reset potential by a programmable conductance for a finite refractory period as soon as the membrane potential crosses a firing threshold $V_\text{th}$.
All neurons are individually configurable via an on-chip analog parameter memory \citep{hock13analogmemory} and a set of digital control values.

\begin{figure}[t]
	\begin{center}
		\scalebox{\FigureScale}{
			\begin{tikzpicture}[>=stealth,x=1.4cm,y=1.4cm]
	\def\columns{2}
	\def\rows{2}
	
	\pgfmathsetseed{1230}

	\def\spikecolors{{"red","blue","orange"}}

	\newcommand{\drawspiketrain}[4]{%
		\pgfmathsetseed{#1}
		\foreach \i in {0,1,...,200} {
			\pgfmathparse{int(random(0, 100)/1.0)}
			\ifnum \numexpr\pgfmathresult > \numexpr100-#2
				\draw[draw=#3] #4 ++ (\i*0.0035,0.05) -- ++(0,0.2);
			\fi
		}
	}

	\begin{scope}
		\foreach \row in {0,1,...,\numexpr\rows-1} {
			\draw[black,thick] (-1.0,-\row) -- (-0.4,-\row);
		}
	\end{scope}

	\begin{scope}
		\begin{scope}
			\foreach[count=\address from 0] \strength/\color in {1/red,1/green,1/blue} {
				\draw[thick] (-2.8,-1 + 0.35*\address-0.35-0.175) -- ++(1.1,0) to[out=0,in=180] (-1.3,-1) -- (-1.0,-1);
				\pgfmathparse{\spikecolors[\address]}
				\colorlet{sc}{\pgfmathresult!100}
				\drawspiketrain{\address+1}{\strength}{sc}{(-2.8,-1 + \address*0.35-0.35-0.175)}

				\draw[draw=sc,fill=white,rounded corners=0.2] (-1.5-0.25-0.25,-1+\address*0.35-0.35+0.05-0.175) rectangle ++(0.2,0.2)
					node[pos=0.5,sc] {\texttt\address};
			}
		\end{scope}
		
		\foreach[count=\address from 0] \strength/\color in {1/red,1/green,1/blue} {
			\pgfmathparse{\spikecolors[\address]}
			\colorlet{sc}{\pgfmathresult!100}
			\drawspiketrain{\address+1}{\strength}{sc}{(-1.2,-1)}
		}
	\end{scope}

	\pgfmathsetseed{1236}
	\begin{scope}
		\foreach \x in {0,1,...,\numexpr\columns-1}
			\draw[thick,black,-stealth] (\x,0.5) -- ++(0,-\rows) -- ++(0.0,-0.35);
		\foreach \y in {0,1,...,\numexpr\rows-1}
			\draw[thick,black] (-0.5,-\y) -- ++(\columns,0);

		\foreach \column in {0,1,...,\numexpr\columns-1} {
			\foreach \row in {0,1,...,\numexpr\rows-1} {
				\pgfmathparse{int(random(0,2))}
				\edef\address{\pgfmathresult}
				\pgfmathparse{\spikecolors[\address]}
				\colorlet{sc}{\pgfmathresult!100}
				
				\ifthenelse{\column = 0 \AND \row = 0} {
					\def\drawlevel{100}
					\def\filllevel{20}
					\def\textlevel{100}
				} {
					\def\drawlevel{20}
					\def\filllevel{5}
					\def\textlevel{50}
				}
				\draw[draw=black!\drawlevel,fill=white,rounded corners=1.0] (\column,\row-1) ++ (-0.2,0.15) rectangle ++(0.4,-0.6);
				\draw[draw=black!\drawlevel] (\column,\row-1) ++ (-0.2,-0.15) -- ++(0.4,0);
				\draw[draw=none] (\column,\row-1) ++ (-0.2,0.15) rectangle ++(0.4,-0.3) node[pos=0.5,color=sc!\textlevel] {\texttt\address};
				\draw[draw=none] (\column,\row-1) ++ (0,-0.45) ++ (-0.2,-0.0) rectangle ++(0.4,0.3)
					node[pos=0.5,color=black!\textlevel] {\pgfmathparse{int(random(0,63))}\texttt\pgfmathresult};

			}
		}
	\end{scope}

	\begin{scope}
		\foreach \neuron in {0,...,\numexpr\columns-1} {
			\ifthenelse{\neuron = 0} {
				\def\drawlevel{100}
				\def\filllevel{20}
				\def\textlevel{100}
			} {
				\def\drawlevel{20}
				\def\filllevel{5}
				\def\textlevel{50}
			}
			
			\draw (\neuron,-\rows-0.4) ++(0.3,-0.1) -- ++(-0.3,0.55) -- ++(-0.3,-0.55) -- cycle;
			\draw (\neuron,-\rows-0.3) node[] {\neuron};

		}
	\end{scope}	


\end{tikzpicture}
		}
	\end{center}
	\caption{
        \tb{Synaptic event filtering enables efficient structural plasticity.}
    	Events are identified with an address denoting their source (numbered and marked by color).
    	Spike trains from different origins can be overlayed and injected into a single synapse row.
    	Synapses filter afferent events by comparing the source address to a label stored in their local \acrshort{sram} and forward only matching spikes to the postsynaptic neurons.
    	Addresses and labels can be reconfigured by the \acrshort{ppu} to implement weight dynamics and structural changes.
    	}
	\label{fig:synapse_array}
\end{figure}
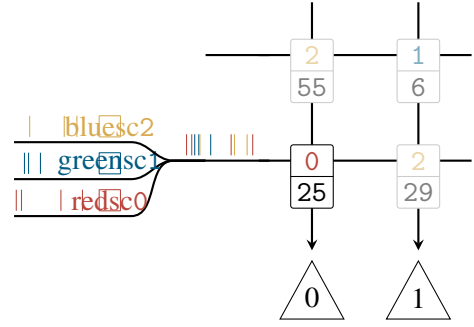

Each neuron is associated with a column of 32 synapse circuits\footnotemark[1] \citep{friedmann2016hybridlearning}, which receive their inputs from the chip's digital backend.
Incoming events are tagged with addresses, which denote their presynaptic origins (Fig.~\ref{fig:synapse_array}).
A \SI{6}{\bit} label is stored alongside the \SI{6}{\bit} weight in the synapse-local \gls{sram}.
It allows to filter afferent spike trains by their addresses; only an event matching the locally stored label is forwarded to the postsynaptic neuron circuit.
Each synapse also implements an analog circuit for measuring pairwise correlations between pre- and post-synaptic spike events \citep{friedmann2016hybridlearning}, enabling access to various forms of learning rules based on nearest-neighbour \gls{stdp}.
The analog correlation traces are made accessible by the \gls{cadc}.

The versatility of the BrainScaleS-2 architecture is substantially augmented by the incorporation of a freely programmable embedded microprocessor \citep{friedmann2016hybridlearning}, which enables the execution of custom programs interacting with the analog emulation of the neuro-synaptic dynamics. 
Together with the \gls{simd} vector unit, which is tightly coupled to the synapse array's \gls{sram} controller and the \gls{cadc}, it forms the \gls{ppu}, which allows the efficient on-chip implementation of synaptic plasticity rules.
Access to the chip-internal configuration bus further allows the processor to also reconfigure all other components of the neuromorphic system during experiment execution.
The \gls{ppu} can thus be used for a vast array of applications such as near-arbitrary learning rules, on-line circuit calibration, or the co-simulation of an environment capable of continuous interaction with the network running on the neuromorphic core.
On the prototype system used in this work, the plasticity processor runs with a frequency of \SI{100}{\mega\hertz}.
Its \gls{simd} unit operates in parallel on slices of 16 synapses%
\footnote[2]{Later versions of the system include multiple \glspl{ppu}, which are clocked at frequencies of up to \SI{400}{\mega\hertz} and feature vector registers capable of handling slices of 128 synapses.}%
.
Iterating row-wise across the synapse matrix, this parallel access lets plasticity algorithms scale $\sim \mathcal{O}(m)$ with the the indegree $m$ but $\sim \mathcal{O}(1)$ with the number of postsynaptic neurons.

A \gls{fpga} is used to interface the \gls{asic} with a host computer.
It also provides sequencing mechanisms for experiment control and spike handling.
Our experiments were based on this paradigm.
However, it was shown that the \gls{ppu} can replace all of the \gls{fpga}'s functionality during experiment runtime \citep{wunderlich2019demonstrating}, dramatically reducing the overall system's power consumption.
In this case, the \gls{fpga} is only used for initial configuration as well as to read out and store observables for later analysis and visualization.
This is an essential prerequisite for the scalability of the BrainScaleS-2 architecture.

\subsection{Pruning and reassignment of synapses}
\label{ssec:pruning}
\begin{figure}[t!]
	\begin{center}
		\scalebox{\FigureScale}{
			\begin{tikzpicture}
				\node[anchor=north west] (plot) at (0.0,0) {
\begingroup%
\makeatletter%
\begin{pgfpicture}%
\pgfpathrectangle{\pgfpointorigin}{\pgfqpoint{1.700000in}{1.200000in}}%
\pgfusepath{use as bounding box, clip}%
\begin{pgfscope}%
\pgfsetbuttcap%
\pgfsetmiterjoin%
\pgfsetlinewidth{0.000000pt}%
\definecolor{currentstroke}{rgb}{0.000000,0.000000,0.000000}%
\pgfsetstrokecolor{currentstroke}%
\pgfsetstrokeopacity{0.000000}%
\pgfsetdash{}{0pt}%
\pgfpathmoveto{\pgfqpoint{0.000000in}{0.000000in}}%
\pgfpathlineto{\pgfqpoint{1.700000in}{0.000000in}}%
\pgfpathlineto{\pgfqpoint{1.700000in}{1.200000in}}%
\pgfpathlineto{\pgfqpoint{0.000000in}{1.200000in}}%
\pgfpathclose%
\pgfusepath{}%
\end{pgfscope}%
\begin{pgfscope}%
\pgfsetbuttcap%
\pgfsetmiterjoin%
\pgfsetlinewidth{0.000000pt}%
\definecolor{currentstroke}{rgb}{0.000000,0.000000,0.000000}%
\pgfsetstrokecolor{currentstroke}%
\pgfsetstrokeopacity{0.000000}%
\pgfsetdash{}{0pt}%
\pgfpathmoveto{\pgfqpoint{0.153000in}{0.120000in}}%
\pgfpathlineto{\pgfqpoint{1.666000in}{0.120000in}}%
\pgfpathlineto{\pgfqpoint{1.666000in}{1.176000in}}%
\pgfpathlineto{\pgfqpoint{0.153000in}{1.176000in}}%
\pgfpathclose%
\pgfusepath{}%
\end{pgfscope}%
\begin{pgfscope}%
\definecolor{textcolor}{rgb}{0.317647,0.317647,0.317647}%
\pgfsetstrokecolor{textcolor}%
\pgfsetfillcolor{textcolor}%
\pgftext[x=0.909500in,y=0.064444in,,top]{\color{textcolor}\rmfamily\fontsize{6.664000}{7.996800}\selectfont time}%
\end{pgfscope}%
\begin{pgfscope}%
\definecolor{textcolor}{rgb}{0.317647,0.317647,0.317647}%
\pgfsetstrokecolor{textcolor}%
\pgfsetfillcolor{textcolor}%
\pgftext[x=0.097444in,y=0.648000in,,bottom,rotate=90.000000]{\color{textcolor}\rmfamily\fontsize{6.664000}{7.996800}\selectfont weight}%
\end{pgfscope}%
\begin{pgfscope}%
\pgfpathrectangle{\pgfqpoint{0.153000in}{0.120000in}}{\pgfqpoint{1.513000in}{1.056000in}}%
\pgfusepath{clip}%
\pgfsetroundcap%
\pgfsetroundjoin%
\pgfsetlinewidth{0.803000pt}%
\definecolor{currentstroke}{rgb}{0.000000,0.000000,0.000000}%
\pgfsetstrokecolor{currentstroke}%
\pgfsetdash{}{0pt}%
\pgfpathmoveto{\pgfqpoint{0.143000in}{0.456041in}}%
\pgfpathlineto{\pgfqpoint{0.243780in}{0.504000in}}%
\pgfpathlineto{\pgfqpoint{0.344647in}{0.360000in}}%
\pgfpathlineto{\pgfqpoint{0.445513in}{0.360000in}}%
\pgfpathlineto{\pgfqpoint{0.546380in}{0.216000in}}%
\pgfusepath{stroke}%
\end{pgfscope}%
\begin{pgfscope}%
\pgfpathrectangle{\pgfqpoint{0.153000in}{0.120000in}}{\pgfqpoint{1.513000in}{1.056000in}}%
\pgfusepath{clip}%
\pgfsetroundcap%
\pgfsetroundjoin%
\pgfsetlinewidth{0.803000pt}%
\definecolor{currentstroke}{rgb}{0.000000,0.000000,0.000000}%
\pgfsetstrokecolor{currentstroke}%
\pgfsetdash{}{0pt}%
\pgfpathmoveto{\pgfqpoint{0.143000in}{0.456000in}}%
\pgfpathlineto{\pgfqpoint{0.243780in}{0.456000in}}%
\pgfpathlineto{\pgfqpoint{0.344647in}{0.552000in}}%
\pgfpathlineto{\pgfqpoint{0.445513in}{0.552000in}}%
\pgfpathlineto{\pgfqpoint{0.546380in}{0.648000in}}%
\pgfpathlineto{\pgfqpoint{0.647247in}{0.696000in}}%
\pgfusepath{stroke}%
\end{pgfscope}%
\begin{pgfscope}%
\pgfpathrectangle{\pgfqpoint{0.153000in}{0.120000in}}{\pgfqpoint{1.513000in}{1.056000in}}%
\pgfusepath{clip}%
\pgfsetroundcap%
\pgfsetroundjoin%
\pgfsetlinewidth{0.803000pt}%
\definecolor{currentstroke}{rgb}{0.000000,0.000000,0.000000}%
\pgfsetstrokecolor{currentstroke}%
\pgfsetdash{}{0pt}%
\pgfpathmoveto{\pgfqpoint{0.143000in}{0.455876in}}%
\pgfpathlineto{\pgfqpoint{0.243780in}{0.312000in}}%
\pgfpathlineto{\pgfqpoint{0.344647in}{0.264000in}}%
\pgfpathlineto{\pgfqpoint{0.445513in}{0.168000in}}%
\pgfpathlineto{\pgfqpoint{0.546380in}{0.168000in}}%
\pgfusepath{stroke}%
\end{pgfscope}%
\begin{pgfscope}%
\pgfpathrectangle{\pgfqpoint{0.153000in}{0.120000in}}{\pgfqpoint{1.513000in}{1.056000in}}%
\pgfusepath{clip}%
\pgfsetroundcap%
\pgfsetroundjoin%
\pgfsetlinewidth{0.803000pt}%
\definecolor{currentstroke}{rgb}{0.000000,0.000000,0.000000}%
\pgfsetstrokecolor{currentstroke}%
\pgfsetdash{}{0pt}%
\pgfpathmoveto{\pgfqpoint{0.647247in}{0.456000in}}%
\pgfpathlineto{\pgfqpoint{0.748113in}{0.408000in}}%
\pgfpathlineto{\pgfqpoint{0.848980in}{0.264000in}}%
\pgfpathlineto{\pgfqpoint{0.949847in}{0.312000in}}%
\pgfpathlineto{\pgfqpoint{1.050713in}{0.360000in}}%
\pgfusepath{stroke}%
\end{pgfscope}%
\begin{pgfscope}%
\pgfpathrectangle{\pgfqpoint{0.153000in}{0.120000in}}{\pgfqpoint{1.513000in}{1.056000in}}%
\pgfusepath{clip}%
\pgfsetroundcap%
\pgfsetroundjoin%
\pgfsetlinewidth{0.803000pt}%
\definecolor{currentstroke}{rgb}{0.000000,0.000000,0.000000}%
\pgfsetstrokecolor{currentstroke}%
\pgfsetdash{}{0pt}%
\pgfpathmoveto{\pgfqpoint{0.647247in}{0.696000in}}%
\pgfpathlineto{\pgfqpoint{0.748113in}{0.840000in}}%
\pgfpathlineto{\pgfqpoint{0.848980in}{0.792000in}}%
\pgfpathlineto{\pgfqpoint{0.949847in}{0.840000in}}%
\pgfpathlineto{\pgfqpoint{1.050713in}{0.888000in}}%
\pgfpathlineto{\pgfqpoint{1.151580in}{1.032000in}}%
\pgfusepath{stroke}%
\end{pgfscope}%
\begin{pgfscope}%
\pgfpathrectangle{\pgfqpoint{0.153000in}{0.120000in}}{\pgfqpoint{1.513000in}{1.056000in}}%
\pgfusepath{clip}%
\pgfsetroundcap%
\pgfsetroundjoin%
\pgfsetlinewidth{0.803000pt}%
\definecolor{currentstroke}{rgb}{0.000000,0.000000,0.000000}%
\pgfsetstrokecolor{currentstroke}%
\pgfsetdash{}{0pt}%
\pgfpathmoveto{\pgfqpoint{0.647247in}{0.456000in}}%
\pgfpathlineto{\pgfqpoint{0.748113in}{0.504000in}}%
\pgfpathlineto{\pgfqpoint{0.848980in}{0.504000in}}%
\pgfpathlineto{\pgfqpoint{0.949847in}{0.600000in}}%
\pgfpathlineto{\pgfqpoint{1.050713in}{0.696000in}}%
\pgfpathlineto{\pgfqpoint{1.151580in}{0.648000in}}%
\pgfusepath{stroke}%
\end{pgfscope}%
\begin{pgfscope}%
\pgfpathrectangle{\pgfqpoint{0.153000in}{0.120000in}}{\pgfqpoint{1.513000in}{1.056000in}}%
\pgfusepath{clip}%
\pgfsetroundcap%
\pgfsetroundjoin%
\pgfsetlinewidth{0.803000pt}%
\definecolor{currentstroke}{rgb}{0.000000,0.000000,0.000000}%
\pgfsetstrokecolor{currentstroke}%
\pgfsetdash{}{0pt}%
\pgfpathmoveto{\pgfqpoint{1.151580in}{0.456000in}}%
\pgfpathlineto{\pgfqpoint{1.252447in}{0.360000in}}%
\pgfpathlineto{\pgfqpoint{1.353313in}{0.408000in}}%
\pgfpathlineto{\pgfqpoint{1.454180in}{0.312000in}}%
\pgfpathlineto{\pgfqpoint{1.555047in}{0.168000in}}%
\pgfusepath{stroke}%
\end{pgfscope}%
\begin{pgfscope}%
\pgfpathrectangle{\pgfqpoint{0.153000in}{0.120000in}}{\pgfqpoint{1.513000in}{1.056000in}}%
\pgfusepath{clip}%
\pgfsetroundcap%
\pgfsetroundjoin%
\pgfsetlinewidth{0.803000pt}%
\definecolor{currentstroke}{rgb}{0.000000,0.000000,0.000000}%
\pgfsetstrokecolor{currentstroke}%
\pgfsetdash{}{0pt}%
\pgfpathmoveto{\pgfqpoint{1.151580in}{1.032000in}}%
\pgfpathlineto{\pgfqpoint{1.252447in}{1.128000in}}%
\pgfpathlineto{\pgfqpoint{1.353313in}{1.032000in}}%
\pgfpathlineto{\pgfqpoint{1.454180in}{1.032000in}}%
\pgfpathlineto{\pgfqpoint{1.555047in}{0.984000in}}%
\pgfpathlineto{\pgfqpoint{1.655913in}{1.032000in}}%
\pgfusepath{stroke}%
\end{pgfscope}%
\begin{pgfscope}%
\pgfpathrectangle{\pgfqpoint{0.153000in}{0.120000in}}{\pgfqpoint{1.513000in}{1.056000in}}%
\pgfusepath{clip}%
\pgfsetroundcap%
\pgfsetroundjoin%
\pgfsetlinewidth{0.803000pt}%
\definecolor{currentstroke}{rgb}{0.000000,0.000000,0.000000}%
\pgfsetstrokecolor{currentstroke}%
\pgfsetdash{}{0pt}%
\pgfpathmoveto{\pgfqpoint{1.151580in}{0.648000in}}%
\pgfpathlineto{\pgfqpoint{1.252447in}{0.696000in}}%
\pgfpathlineto{\pgfqpoint{1.353313in}{0.744000in}}%
\pgfpathlineto{\pgfqpoint{1.454180in}{0.696000in}}%
\pgfpathlineto{\pgfqpoint{1.555047in}{0.744000in}}%
\pgfpathlineto{\pgfqpoint{1.655913in}{0.792000in}}%
\pgfusepath{stroke}%
\end{pgfscope}%
\begin{pgfscope}%
\pgfpathrectangle{\pgfqpoint{0.153000in}{0.120000in}}{\pgfqpoint{1.513000in}{1.056000in}}%
\pgfusepath{clip}%
\pgfsetbuttcap%
\pgfsetroundjoin%
\pgfsetlinewidth{0.803000pt}%
\definecolor{currentstroke}{rgb}{0.000000,0.000000,0.000000}%
\pgfsetstrokecolor{currentstroke}%
\pgfsetstrokeopacity{0.400000}%
\pgfsetdash{{4.000000pt}{4.000000pt}}{0.000000pt}%
\pgfpathmoveto{\pgfqpoint{0.153000in}{0.600000in}}%
\pgfpathlineto{\pgfqpoint{1.666000in}{0.600000in}}%
\pgfusepath{stroke}%
\end{pgfscope}%
\begin{pgfscope}%
\pgfsetrectcap%
\pgfsetmiterjoin%
\pgfsetlinewidth{0.501875pt}%
\definecolor{currentstroke}{rgb}{0.317647,0.317647,0.317647}%
\pgfsetstrokecolor{currentstroke}%
\pgfsetdash{}{0pt}%
\pgfpathmoveto{\pgfqpoint{0.153000in}{0.120000in}}%
\pgfpathlineto{\pgfqpoint{0.153000in}{1.176000in}}%
\pgfusepath{stroke}%
\end{pgfscope}%
\begin{pgfscope}%
\pgfsetrectcap%
\pgfsetmiterjoin%
\pgfsetlinewidth{0.501875pt}%
\definecolor{currentstroke}{rgb}{0.317647,0.317647,0.317647}%
\pgfsetstrokecolor{currentstroke}%
\pgfsetdash{}{0pt}%
\pgfpathmoveto{\pgfqpoint{0.153000in}{0.120000in}}%
\pgfpathlineto{\pgfqpoint{1.666000in}{0.120000in}}%
\pgfusepath{stroke}%
\end{pgfscope}%
\end{pgfpicture}%
\makeatother%
\endgroup%
};

				\draw[green,line width=1pt,-stealth] (plot.south west) ++ (4.33,2.73) ++ (0.0,0.07) -- ++(0.0, 0.35) node[pos=0.35,right,xshift=3pt]
					{\footnotesize Hebbian potentiation};
				\draw[red,line width=1pt,-stealth] (plot.south west) ++ (4.33,2.73) ++ (0.0,-0.07) -- ++(0.0,-0.35) node[pos=0.35,right,xshift=3pt]
					{\footnotesize regularization};
				
				\draw[gray] (plot.south west) ++ (4.20,1.62) node[right,xshift=3pt] {\footnotesize $\theta_\text{w}$};

				\draw[orange,line width=1pt] (plot.south west) ++ (2.93,1.17) circle (9pt);
				\draw[orange,line width=1pt] (plot.south west) ++ (2.93,1.17) ++ (-10:11pt) -- ++(-10:40pt) node [pos=1.0,right,xshift=3pt]
					{\footnotesize pruning and reassignment};
			\end{tikzpicture}
		}
	\end{center}
	\caption{\tb{Illustration of weight dynamics.}
	The evolution of synaptic weights is governed by a Hebbian potentiation term and a regularizing force of opposing sign.
	A stochastic component in the weight update term leads to a random walk.
	Synapses with an efficacy below the pruning threshold $\thetaw$ are regularly reassigned to new receptors, allowing neurons to find more informative presynaptic partners, to which the connections can then be strengthened.
	}
	\label{fig:weight_dynamics}
\end{figure}
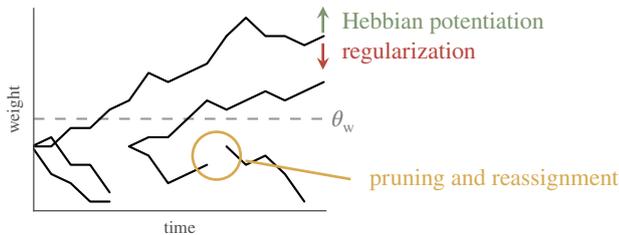

We propose a mechanism and an optimized hardware implementation for structural plasticity inspired by two well-established biological observations.
First, we assume that important, informative synapses have larger absolute weights.
In our particular setting, this is achieved by Hebbian learning, augmented by slow unlearning, as outlined in Section \ref{ssec:updates}, but this assumption holds for many other plasticity mechanisms as well \citep{oja1982simplified,urbanczik2014learning,fremaux2016neuromodulated,mostafa2017supervised,zenke2018superspike}.
Second, we enable the network to manage its limited synaptic resources towards potentially improving its performance by removing weak synapses and creating new ones instead.

A synapse's eligibility for pruning is determined by the value of its weight:
it is removed in case its efficacy falls below a threshold $\thetaw$ (Fig.~\ref{fig:weight_dynamics}).
Whenever an afferent synapse is removed, the postsynaptic neuron replaces it with a connection to a randomly selected presynaptic partner, thus conserving its indegree.
The newly created synapse is intialized with a low weight $\winit$.
The pruning process takes place at a slower timescale than the network dynamics and weight updates, giving the synaptic weights time to develop and integrate over multiple update periods.

The implementation on BrainScaleS-2 exploits an in-synapse resolution of the connectome.
Each event carries a label denoting its origin, allowing synapses to distinguish different sources.
A synapse filters afferent spike trains by comparing this event address to the locally stored value and forwards only matching events to its postsynaptic neuron.
Pruning and reassigning of synapses is implemented by remapping the label stored in the synapse-local \gls{sram}, which effectively eliminates the previous connection.

\begin{algorithm}[t!]
	\small
	\begin{algorithmic}
			\FOR{\texttt{row in 0 ... 31}}
				\STATE \texttt{w $\leftarrow$ synram\_weights\_read(row)}\;
				\STATE \texttt{w $\leftarrow$ w + alpha * min(f\_max,correlation\_read(row))}\;
				\STATE \texttt{w $\leftarrow$ w - beta * w * rates\_read()}\;
				\STATE \texttt{w $\leftarrow$ w + gamma * rng(-1,1)}\;
				\IF{{\texttt w} $<$ {\texttt theta\_w}}
					\STATE {\texttt w} $\leftarrow$ \texttt{w\_init}\;
					\STATE {\texttt a} $\leftarrow$ \texttt{rng(0,k)}\;
					\STATE \texttt{synram\_labels\_write(row,a)}\;
				\ENDIF
				\STATE \texttt{synram\_weights\_write(row,w)}\;
			\ENDFOR
	\end{algorithmic}
	\caption{
		\tb{Plasticity algorithm including weight updates and structural reconfiguration.}
		The update algorithm is executed by the on-chip \gls{ppu} and is applied iteratively to the synapse rows.
		Synapses within a row are processed in parallel.
		The \acrshort{ppu} supports \acrshort{simd} vector instructions including arithmetic operations and access to the synaptic memory (\texttt{synram\_weights\_\{read,write\}()}, \texttt{synram\_labels\_write()}) and \acrshort{cadc} data (\texttt{correlation\_read()}).
		It has also access to the neuronal firing rates (\texttt{rates\_read()}) and uniform pseudo-random number generators (\texttt{rng()}).
		}
	\label{alg:plasticity_algorithm}
\end{algorithm}

As compared to other synaptic pruning and reassignment strategies, our algorithm and implementation of structural plasticity requires a particularly low overhead.
The in-synapse definition of the connectome allows a purely local reassignment mechanism which avoids global access patterns;
for example, no reordering of routing tables is required, which can otherwise lead to increased computational complexity \citep{liu2018memory}.
At its core, reassignment only involves a single \gls{sram} access.
Also the evaluation of the pruning condition and the selection of a new presynaptic partner can be realized with just a few simple instructions (Fig.~\ref{alg:plasticity_algorithm}).

\subsection{Correlation-driven weight update algorithm}
\label{ssec:updates}
The synaptic reassignment algorithm described above 
is based on the assumption that high weights are assigned to informative synapses, which emerges through a manifold of learning strategies.
In this work, we chose Hebbian weight dynamics, as on BrainScaleS-2 they allow a fully local implementation of weight updates and, furthermore, can be extended to form more advanced learning rules~\citep{fremaux2016neuromodulated,neftci2014event}.
Here,
the temporal evolution of the synaptic weights $w_{ij}$, which is illustrated in Fig.~\ref{fig:weight_dynamics}, obeys the following equations:
\begin{align}
	\Delta w_{ij} &= \alpha \cdot f(S_i, S_j) - \beta \cdot \nu_{i} w_{ij} + \gamma \cdot \eta_{ij} \; , \label{eq:delta_w} \\
	f(S_i, S_j) &= \min \left[ \fmax , \sum_k \exp \left( - \frac{t_i^k - \max_l \left[ t_j^l < t_i^k \right]}{\taustdp} \right) \right] \; . \label{eq:stdp}
\end{align}
The update rule (Eqn.~\ref{eq:delta_w}) consists of three terms.
The first term represents an implementation of \gls{stdp} and depends on the post- and presynaptic spiketrains $S_i$ and $S_j$, defined as vectors of ordered spike times $t_i^k$ and $t_j^l$.
The STDP kernel is exponential and positive for causal presynaptic spikes and zero for anti-causal ones, with a cutoff at a maximum value $\fmax$ (Eqn.~\ref{eq:stdp}).
The second term implements homeostasis (by penalizing large postsynaptic firing rates) and forgetting (as an exponential decay).
This regularizer encourages competition between the afferent synapses of a neuron.
The third term induces exploration by means of a uniformly drawn random variable $\eta_{ij}$ leading to an unbiased random walk, similar to the work by~\citet{kappel2015synaptic}.
On BrainScaleS-2, this stochastic component helps to overcome local minima induced by analog fixed-pattern noise and integer arithmetics.
The three components are weighted with positive factors $\alpha$, $\beta$, and $\gamma$, respectively.

All three contributions to the weight update rule can be mapped to specialized hardware components.
The \gls{stdp}-derived term is based on correlation traces.
These observables are measured in analog synapse-local circuits and then digitized using the \gls{cadc} (section~\ref{ssec:arch}).

As stated above, the correlation values are capped.
This is required to reduce the imbalance introduced by fixed-pattern deviations in the correlation measurement circuits' sensitivity, as some of these analog sensors might systematically detect stronger correlation values than others.
This can lead to an overly strong synchronisation of the respective receptor and label neurons, in turn resulting in a self-amplifying potentiation of the corresponding weight and a resulting dominance over the teacher spike train.
In principle, a decrease of $\alpha$ could dampen such feedback, but the corresponding reduction of the exponential \gls{stdp} kernel can be difficult to reconcile with fixed-point calculations of limited precision.

The homeostatic component requires access to the postsynaptic firing rates, which are read from spike counters via the on-chip configuration bus.
Stochasticity is provided by an \emph{xorshift} algorithm \citep{marsaglia2003xorshift} implemented in software.\footnote[3]{Later versions of the system feature hardware acceleration for the generation of pseudo-random numbers.}
The individual contributions are processed and accumulated on the embedded processor:
using the \gls{simd} vector unit, it is able to handle slices of 16 synapses in parallel.

\subsection{Classification task}
\label{ssec:task}
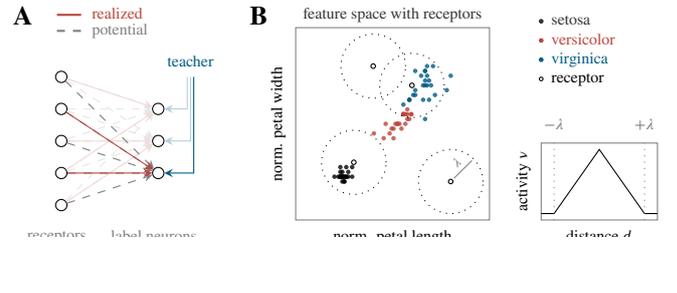
\begin{figure}[t]
	\begin{center}
		\scalebox{\FigureScale}{
			\begin{tikzpicture}
				\draw[use as bounding box,draw=none] (-0.1,0.0) rectangle ++(8.8,-3.3);

				\node[anchor=north west] (network) at (0.0,0) {\pgfmathsetseed{123456789}%
\tikzset{expressed/.style={-stealth,shorten <=4pt,shorten >=6pt,ultra thick}}%
\tikzset{potential/.style={-stealth,shorten <=4pt,shorten >=6pt,very thick,dotted}}%
\usetikzlibrary{calc}
\begin{tikzpicture}[scale=0.85]
	\node[below,gray] at (0.0,-0.3) {\tiny receptors};
	\node[below,gray] at (1.5,-0.3) {\tiny label neurons};
	
	\foreach \i in {0,...,4} \node[draw,circle,inner sep=1.5pt] (input\i) at (0,\the\numexpr0.5*\i) {};
	\foreach \i in {0,...,2} \node[draw,circle,inner sep=1.5pt] (output\i) at (1.5,\the\numexpr0.5*\i+0.5) {};

	\foreach \i in {0,...,4}
		\foreach \j in {0,...,2} \draw[-stealth,gray!20!white,dashed] (input\i) -- (output\j);

	\draw[-stealth,red!20!white] (input0) -- (output1);
	\draw[-stealth,red!20!white] (input2) -- (output1);

	\draw[-stealth,red!20!white] (input1) -- (output2);
	\draw[-stealth,red!20!white] (input4) -- (output2);
	
	\foreach \i in {0,...,4} \draw[-stealth,gray!100!white,dashed] (input\i) -- (output0);

	\draw[-stealth,red!100!white] (input1) -- (output0);
	\draw[-stealth,red!100!white] (input3) -- (output0);

	\draw[stealth-,blue!100!white] (output0) -- ++(0.55,0) -- ++(0,1.5);
	\draw[stealth-,blue!30!white] (output1) -- ++(0.50,0) -- ++(0,1.0);
	\draw[stealth-,blue!30!white] (output2) -- ++(0.45,0) -- ++(0,0.5);

	\draw (output1) ++ (0.5,1.0) node[blue,above] {\tiny teacher};

	\draw[thick,red] (0.0,2.90) -- ++(0.375,0.0) node[right] {\tiny \vphantom{gt}realized};
	\draw[thick,gray,dashed] (0.0,2.65) -- ++(0.375,0.0) node[right] {\tiny \vphantom{gt}potential};
\end{tikzpicture}};
				\node[anchor=north west] (feature) at (3.3,0.0) {\pgfmathsetseed{123456789}%
\tikzset{expressed/.style={-stealth,shorten <=4pt,shorten >=6pt,ultra thick}}%
\tikzset{potential/.style={-stealth,shorten <=4pt,shorten >=6pt,very thick,dotted}}%
\pgfplotstableread[header=false]{\relativepath iris.csv}{\iris}%
\begin{tikzpicture}[scale=0.85]

	\def\scale{3}

	\begin{scope}[yshift=0,every node/.append style={yslant=0.0,xslant=0.0},yslant=0.0,xslant=0.0,yscale=1.0]
		\draw[gray] (0,0) rectangle ++(\scale,\scale);
		\foreach \i in {0,2,...,149} {
			\pgfplotstablegetelem{\number\numexpr\i}{[index]0}\of\iris
			\let\x\pgfplotsretval
			\pgfplotstablegetelem{\number\numexpr\i}{[index]1}\of\iris
			\let\y\pgfplotsretval
			\pgfplotstablegetelem{\number\numexpr\i}{[index]2}\of\iris
			\let\c\pgfplotsretval

			\fill[fill=\c,opacity=0.8] (\scale*\x,\scale*\y) circle (1pt);
		}

		\coordinate (receptor0) at (0.8*\scale,0.2*\scale);
		\coordinate (receptor1) at (0.3*\scale,0.3*\scale);
		\coordinate (receptor2) at (0.6*\scale,0.7*\scale);
		\coordinate (receptor3) at (0.4*\scale,0.8*\scale);

		\draw[dotted] (receptor0) circle (0.5cm);
		\draw[dotted] (receptor1) circle (0.5cm);
		\draw[dotted] (receptor2) circle (0.5cm);
		\draw[dotted] (receptor3) circle (0.5cm);
		\draw[solid] (receptor0) circle (1pt);
		\draw[solid] (receptor1) circle (1pt);
		\draw[solid] (receptor2) circle (1pt);
		\draw[solid] (receptor3) circle (1pt);

		\draw[gray] (receptor0) ++ (45:2pt) -- ++(45:0.5cm-3pt) node[pos=0.5,above,rotate=45,yshift=-2pt,font=\fontsize{5}{0.5}\selectfont] {$\lambda$};
	\end{scope}
	
	\draw[black!80,fill] (0.5*\scale,3.2) node[anchor=center] {\tiny \vphantom{gt}feature space with receptors};

	\node[below] at (0.5*\scale,0) {\tiny\vphantom{gl}norm. petal length};
	\node[above,rotate=90] at (0,0.5*\scale) {\tiny\vphantom{gl}norm. petal width};
	
	\draw[black!80,fill] (3.8,3.1) circle (0.8pt) node[right] {\tiny \vphantom{gt}setosa};
	\draw[red,fill]      (3.8,2.8) circle (0.8pt) node[right] {\tiny \vphantom{gt}versicolor};
	\draw[blue,fill]     (3.8,2.5) circle (0.8pt) node[right] {\tiny \vphantom{gt}virginica};
	\draw[solid]         (3.8,2.2) circle (0.8pt) node[right] {\tiny \vphantom{gt}receptor};
		
	\draw[gray] (3.8,0) rectangle ++(1.8,1.2);
	\draw[] (3.8,0) ++ (0.0,0.1) -- ++(0.2,0.0) -- ++(0.7,1.0) -- ++(0.7,-1.0) -- ++(0.2,0.0);
	\draw[gray,dotted] (3.8,0) ++ (0.2,0.0) -- ++(0.0,1.2) node[above] {\tiny\vphantom{gt}$-\lambda$};
	\draw[gray,dotted] (3.8,0) ++ (1.6,0.0) -- ++(0.0,1.2) node[above] {\tiny\vphantom{gt}$+\lambda$};

	\node[below] at ($(3.8,0) + (0.9,0.0)$) {\tiny\vphantom{gl}distance $d$};
	\node[above,rotate=90] at ($(3.8,0) + (0.0,0.6)$) {\tiny\vphantom{gl}activity $\nu$};
\end{tikzpicture}};

				\node at ($(network.north west) + ( 0.30,-0.36)$) {\small\bfseries A};
				\node at ($(feature.north west) + ( 0.10,-0.36)$) {\small\bfseries B};
			\end{tikzpicture}
		}
	\end{center}
	\caption{\tb{Sparse network architecture and input encoding.}
	\tb{(A)} The two-layer network consists of a group of receptors and a label population.
	One teacher per label neuron ensures excitation of the correct labels during learning.
	The inputs project onto the label layer with a \emph{potential} all-to-all connectivity (gray), but only a subset of synapses is realized (blue).
	\tb{(B)}
	The receptors are uniformly distributed on the two-dimensional feature space, which is spanned by the petal widths and lengths of Iris flowers belonging to the three classes setosa, versicolor, and virginica.
	A receptor's activity is calculated from its Euclidean distance to a data point according to a triangular kernel with radius $\lambda$.
	}
	\label{fig:network}
\end{figure}

We applied the presented plasticity mechanism including structural reconfiguration to a two-layer network trained to perform a classification task.
The network consisted of a group of spike sources in a receptor layer and a set of label neurons.
These layers were set up such that every postsynaptic neuron could potentially receive input from any presynaptic partner in the receptor layer.
Only a fixed fraction of these potential synapses was expressed at each point in time; the others were dormant, resulting in a sparse connectome.
In addition to the feed-forward connections, label neurons were stimulated by teacher spike sources.
These supervisory projections ensured excitation of a label neuron when an input belonging to their respective class was presented.

The network was trained on the Iris dataset \citep{fisher1936use}.
\citet{schmuker2014neuromorphic} already trained an early predecessor of the BrainScaleS-2 system on the same data, but used an off-chip preprocessing scheme based on principal component analysis to determine optimal receptor locations and static receptive fields.
Here, we reduced the four-dimensional dataset to only two dimensions by selecting petal widths and lengths, renormalized to values between 0.2 and 0.8.
The resulting two-dimensional feature space is shown in Fig.~\ref{fig:network}~B.
On this plane, $n$ virtual receptors were placed at random locations drawn from a uniform distribution.
These receptor neurons emitted Poisson-distributed spike trains with an instantaneous rate determined by their respective Euclidean distances $d$ to a presented data point.
The firing rate was calculated according to a triangular kernel $\nu_i(d) = \hat\nu \cdot \mathrm{max}(0, 1 - d/\lambda)$, with $\hat\nu = \SI{50}{\kilo\hertz}$.
This corresponds to a biologically plausible firing rate of \SI{50}{\hertz}, when taking the system's speedup into account.
The radius $\lambda$ of the receptors was scaled inversely with $\sqrt{n}$ to ensure a reasonable converage of the feature space.

\begin{figure}[t!]
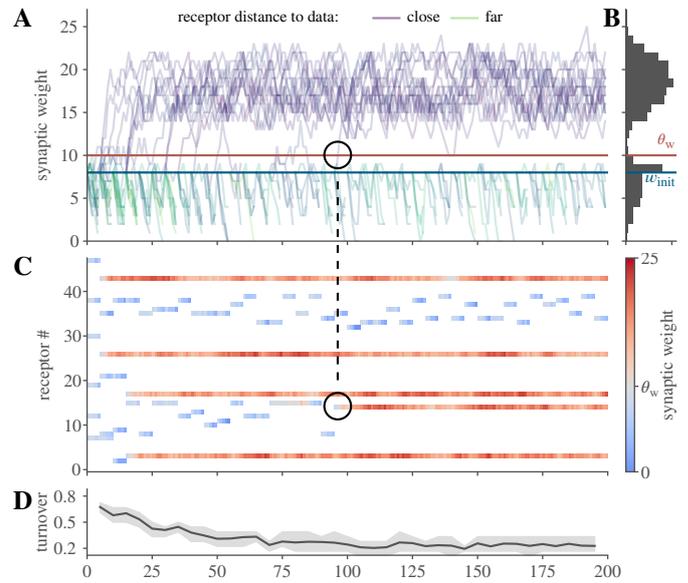

	\begin{center}
		\scalebox{\FigureScale}{
			\begin{tikzpicture}
				\node[anchor=north west] (a) at (0.0, 0.0) {\input{figures/evolution_weights_paper.pgf}};
				\node[anchor=north west] (b) at (7.8, 0.0) {\input{figures/evolution_histogram_paper.pgf}};
				\node[anchor=north west] (c) at (0.0,-3.3) {\input{figures/evolution_mesh_paper.pgf}};
				\node[anchor=north west] (d) at (0.0,-6.4) {\input{figures/evolution_turnover_paper.pgf}};
				
				\node[anchor=north west] at (7.8,-3.3) {\input{figures/evolution_mesh_colorbar_paper.pgf}};

				\node at ($(a.north west) + (0.08,-0.30)$) {\small\bfseries A};
				\node at ($(b.north west) + (0.03,-0.30)$) {\small\bfseries B};
				\node at ($(c.north west) + (0.08,-0.30)$) {\small\bfseries C};
				\node at ($(d.north west) + (0.08,-0.30)$) {\small\bfseries D};

				\draw[thick] (4.22,-5.46) circle (5pt);
				\draw[thick] (4.22,-2.13) circle (5pt);
				\draw[thick,dashed] (4.22,-2.13) ++ (0.0,-0.35) -- ++(0.0,-2.70);
			\end{tikzpicture}
		}
	\end{center}
	\caption{
        \tb{Informative synapses emerge during training.}
	\tb{(A)} Exemplary evolution of realized afferent weights of the ``setosa'' label neuron during the course of a single experiment.
               The line color is determined by the average feature-space distance between the respective receptor and all ``setosa'' data points.
               Synapses that receive inputs from relevant receptors (i.e., those lying close to the features that are relevant for their postsynaptic label neuron) are strengthened towards values that lie above the pruning threshold $\thetaw$.
               All other, less informative synapses remain below $\thetaw$ and are pruned at regular intervals of five epochs.
               For each pruned synapse, a new one is initialized at $\winit$, between the same label neuron and a previously unconnected receptor.
	    \tb{(B)} Distribution of synaptic weights during the last 50 epochs over 20 randomly initialized runs.
               Note that the histogram only takes into account realized synapses, which at all times are only 18 out of 144 potential ones.
	    \tb{(C)} Exemplary evolution of all synaptic weights between the receptor population and the ``setosa'' label neuron.
	           At all times, only $n/k=6$ synapses are realized.
               The transition from blue to red marks the pruning threshold $\thetaw$.
               Note how gray/blue (subthreshold) and white (non-existent) states alternate, marking the pruning of weak synapses and re-initialization of new ones.
	       One of these reassignments is highlighted and referenced to the corresponding threshold crossing in pane A.
	    \tb{(D)} Evolution of the turnover rate (fraction of pruned synapses per epoch) for the 20 runs.
               The solid line marks the mean and the gray area represent the 20 and 80 percentiles.
               As time progresses, the turnover rate converges to approximately \SI{20}{\percent}, indicating that all relevant receptors (on average five) have been found.
               The remaining ``free'' synapses (on average one) keep switching between all other receptors, but are pruned regulary as they are not informative for the respective class.
        }
	\label{fig:weight_distribution}
\end{figure}

\begin{figure*}[t!]
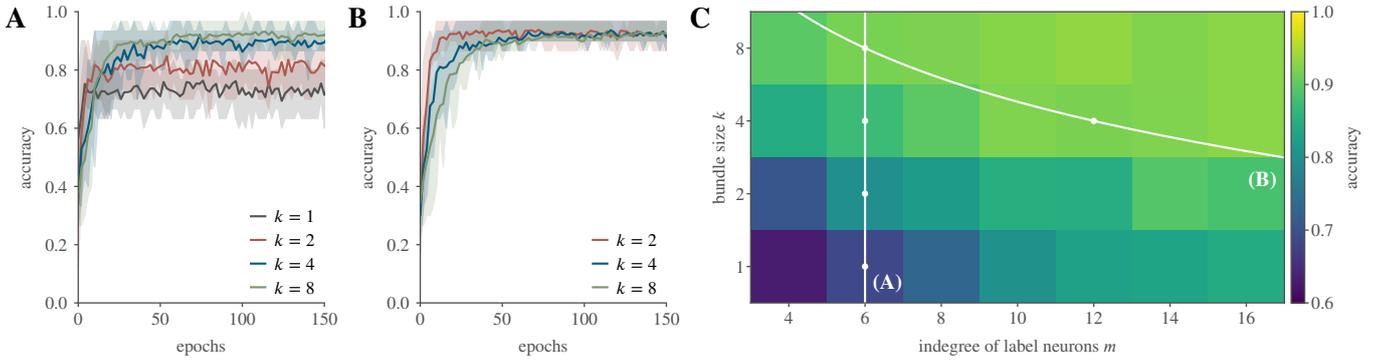

	\begin{center}
		\scalebox{\FigureScale}{
			\begin{tikzpicture}
				\node[anchor=north west] at (0,0) {\input{figures/overload_sweep_paper.pgf}};
				\node[anchor=north west] at (4.5,0) {\input{figures/address_sweep_paper.pgf}};
				\node[anchor=north west] at (9.0,0) {\input{figures/receptor_sweep_paper.pgf}};
				\node at ($(0.0,0) + (0.18,-0.45)$) {\bfseries A};
				\node at ($(4.5,0) + (0.18,-0.45)$) {\bfseries B};
				\node at ($(9.0,0) + (0.18,-0.45)$) {\bfseries C};
			\end{tikzpicture}
		}
    \end{center}
	\caption{
        \tb{Structural plasticity improves learning in sparse networks.}
		\tb{(A)}
            	For a constant indegree $m$ of the label neurons (equivalent with the number of synapse rows on the hardware), classification accuracy improves with larger $k$, as the neurons gain access to an increasing number of receptors $n=km$.
		\tb{(B)}
            	For a constant number of receptors $n$, structural plasticity can compensate for increased sparsity (reduced indegree $m$ induced by a larger bundle size $k$) up to a certain degree.
		\tb{(C)}
			Panels A and B can be embedded into a more extensive sweep over the number of the indegree $m$ and bundle size $k$.
		}
	\label{fig:sweep}
\end{figure*}

To impose a certain level of sparsity, we used the following procedure.
Receptors were randomly grouped into $m$ disjoint bundles of size $k$ and each bundle was injected into a single synapse row.
Within a bundle, each receptor was assigned a unique address.
The sparsity of the connectome, defined as the ratio between the number of unrealized synapses and the number of potential synapses, was thus set to $1 - 1/k = 1 - m/n$.
This setup allowed two degrees of freedom in the control of network sparsity (Fig.~\ref{fig:sweep}).
Increasing the number of receptors $n$ for a fixed synapse count $m$ increased the bundle size $k$ and thus the sparsity as well.
On the other hand, for constant sparsity $1-1/k$, reducing the synapse count $m$ incurred a reduction of the receptor count $n$.
Disjoint bundles do impose restrictions on the realizable connectomes:
First, it is not possible to project to a neuron from two presynaptic sources that both carry relevant information but by chance reside on the same bundle.
Second, multapses, i.e., multiple projections from the same afferent to a single postsynaptic neuron, can not be realized.
These constraints can be circumvented by choosing overlapping bundles or injecting a single bundle into multiple synaptic rows.

The dataset, containing a total of 150 data points, was randomly divided into 120 training and 30 test samples to allow cross validation.
Samples were presented to the network in random order.
For each presented data point, the network's state was determined by a winner-take-all mechanism implemented in software, which compared the firing rates of the label neurons.
Synaptic weights were updated according to Eqn.~\ref{eq:delta_w} after each epoch.
The pruning condition was evaluated regularly every five epochs.

\section{Results}
\begin{figure}[t]
	\begin{center}
		\scalebox{\FigureScale}{
			\begin{tikzpicture}
				\node[anchor=north west] (receptive) at (0.0,0) {\input{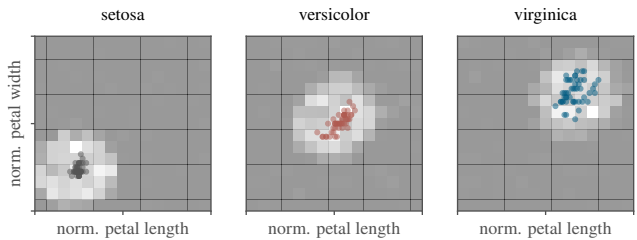}};
			\end{tikzpicture}
		}
	\end{center}
	\caption{
		\tb{Self-organized formation of receptive fields.}
	    	The probability of synapse expression depends on the location of receptors in the feature space and the class of label neurons.
		Each square is shaded according to the probability for a label neuron to have formed a synapse with a receptor lying within that area (lighter for higher probability); estimated from the state at the end of training in 100 experiments with random initial conditions.
		The size of the three emerging clusters is determined by the receptor radius $\lambda$.
	   }
	\label{fig:receptive_fields}
\end{figure}

In this section, we describe experimental results of learning on the BrainScaleS-2 prototype using the plasticity rule and classification task outlined above.
We evaluated the network's performance under varied sparsity constraints and performed sweeps on the hyperparameters to study the robustness of the learning algorithm and demonstrate its efficient use of limited synaptic resources.
Moreover, we highlight the speed of our structural plasticity algorithm, especially in conjunction with its implementation on the BrainScaleS-2 system.

\subsection{Self-configuring receptive fields}

Depending on the nature of the data to be learned, i.e., the distribution of data points in the feature space, some receptors can be more informative than others (Fig.~\ref{fig:network}~B).
Our learning rule naturally selects the most informative receptors, thereby creating a topological order of the label neurons' receptive fields.
This clustering of receptors is driven by the synaptic weight evolution as described by Eqn.~\ref{eq:delta_w} (Fig.~\ref{fig:weight_dynamics}).

Fig.~\ref{fig:weight_distribution} shows this evolution during the course of an experiment.
Starting from their initial values, synapses that contributed causally to the firing of their postsynaptic neurons were potentiated.
After escaping the pruning threshold, they continued evolving until reaching an equilibrium with the homeostatic force.
Weaker connections were regularly pruned and reassigned; the common intialization value manifests itself in a strongly pronounced peak.

The turnover rate, defined as the fraction of pruned synapses, also reflects the formation of receptive fields.
As the receptors were randomly intialized at the beginning of the experiment, they did not reflect the spatial distribution of the dataset.
This resulted in frequent pruning, indicated by a high turnover rate.
Over time, a set of stable synapses was formed and the turnover rate gradually decreased.

The topology of the emergent connectome can be reconstructed from the synaptic labels.
By repeating the experiment with varying seeds and therefore initial conditions, it is possible to calculate a probability density for a synapse to be expressed at a given point on the feature plane.
This map closely resembles the distribution of the presented data (Fig.~\ref{fig:receptive_fields}):
the receptive fields of the respective label neurons cluster around the corresponding samples.
The radius of these clusters is determined by the spread of the data as well as the support and shape of the receptors' kernels.

\subsection{Increased network performance with structural plasticity}

During the course of the training phase, the network's performance was repeatedly evaluated by presenting the test data to the receptor layer.
In this phase, the network's weights and connectome were frozen by disabling weight updates and structural modifications.
To test the network's ability to generalize and reduce the impact of specific positioning of receptors or initial conditions, we trained and evaluated the network starting from 20 randomly drawn initial states.

The evolution of the network's accuracy can be observed in Fig.~\ref{fig:sweep}~A.
Starting from approximately chance level, the performance increased during training and converged to a stable value.
In this specific experiment, we swept the bundle size $k$ while keeping the number of utilized synapse rows $m$ constant, resulting in a variable number of receptors $n = k\cdot m$.
This corresponds to a scenario where the limited afferent synaptic resources per neuron are fully utilized and structural plasticity is required to expand the number of virtual presynaptic partners.
For $k=1$ the network was trained without structural reconfiguration and only had access to a small pool of receptors, resulting in a correspondingly low performance.
As more receptors became available, the classification accuracy increased as well, up to \SI{92.3}{\%} for structural plasticity with a bundle size of 8.

In a second sweep we kept the number of receptors $n$ constant and varied the bundle size $k$.
This resulted in a variable number of realized synapses $m$ and hence different levels of sparsity.
The classification accuracy's evolution for $k \in \{2,4,8\}$ is shown in Fig.~\ref{fig:sweep}~B.
The network achieved a performance of approximately \SI{92}{\%} for all of the sparsity levels.
In this experiment, we showed that learning with structural plasticity allows to reduce the utilization of synaptic resources while conserving the overall network performance.
These results demonstrate that our learning algorithm enables a parsimonious utilization of hardware resources.
The resulting pool of ``free'' synapses can then be used for other purposes, such as for the realization of deeper network structures.
For larger receptor pools, we also note that learning converges more slowly, as the label neurons need more time to explore their respective receptive fields (Fig.~\ref{fig:sweep}~A,B).


Both of the aformentioned experiments can be embedded into a more extensive sweep over receptor counts $n$ and bundle sizes $k$.
In Fig.~\ref{fig:sweep}~C, the two experiments correspond to the two highlighted lines.
Classification performance primarily depends on the count of available receptors -- and to a much lesser extent on the amount of utilized hardware resources.
For the employed classification task, only six synapses were sufficient to reach levels of accuracy otherwise only tangible with more than 32 synapses.

\begin{figure}[t!]
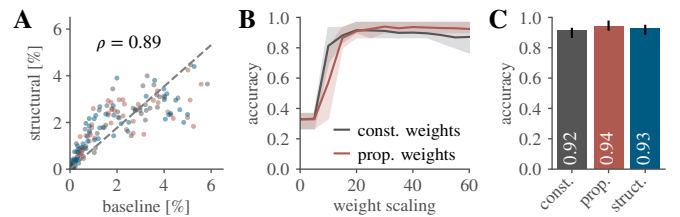

	\begin{center}
		\scalebox{\FigureScale}{
			\begin{tikzpicture}
				\node[anchor=north west] (a) at (0,0) {\input{figures/correlation_paper.pgf}};
				\node[anchor=north west] (b) at (3.0,0) {\input{figures/baseline_sweep_paper.pgf}};
				\node[anchor=north west] (c) at (6.3,0) {\input{figures/baseline_bar_paper.pgf}};
				\node at ($(a.north west) + (0.16,-0.30)$) {\small\bfseries A};
				\node at ($(b.north west) + (0.11,-0.30)$) {\small\bfseries B};
				\node at ($(c.north west) + (0.11,-0.30)$) {\small\bfseries C};
			\end{tikzpicture}
		}
	\end{center}
	\caption{%
		\tb{Comparison of structural plasticity to a baseline estimate.}
		 Structural plasticity yields an accuracy comparable to the one of a network obtained by artificially choosing the most active receptors for each class.
		\tb{(A)} A clear correlation between the synapse expression probabilities of the trained and the baseline networks can be observed.
		\tb{(B)} The network was evaluated for the same set of synapses, but with a variable weight scaling factor.
		In one case the weights were configured homogeneously to a constant value, in the other they were additionally scaled with the receptors' mean activations.
		\tb{(C)} Classification performance for structural plasticity is on par with the respective maxima from B, where the proportional scaling outperforms the constant one.
		}
	\label{fig:baseline}
\end{figure}

We established a baseline accuracy for this task by evaluating the network with artificially set up connectomes.
For each bundle ($k = 8$, $m = 6$) and label neuron, we selected the receptor with the highest mean firing rate for all training data of the respective class.
Analyzing the resulting receptive fields for multiple seeds and plotting the expression probabilities against the ones obtained through structural plasticity (cf. Fig.~\ref{fig:receptive_fields}) shows a clear correlation (Fig.~\ref{fig:baseline}A).
This indicates that the presented structural plasticity algorithm indeed establishes informative synapses.
Due to the threshold-based pruning, which is essential for convergence to a stable connectome, receptors with the highest expression probabilities for the baseline selection were slightly underrepresented in the learnt structure.
We estimated the baseline performance by considering two methods of assigning synaptic weights:
First, we applied the same, constant weight value $w_{ij} = s$ for all established synapses, representing the case of pure structural plasticity and ignoring any weight dynamics.
Second, we chose the weights as $w_{ij} = s \cdot \nu_j/\operatorname{max}_j(\nu_j)$, with the average receptor firing rate $\nu_j$, demonstrating the combined effects of structural reconfiguration and Hebbian learning.
We evaluated both cases with a varied scaling $s$ (Fig.~\ref{fig:baseline}B) and selected the respective maxima for the performance comparison (Fig.~\ref{fig:baseline}C).
Compared to these two baseline measures, which relied on global knowledge of the whole dataset and all receptor activities, the connectome that emerged through structural plasticity, only based on local information, yielded a comparable performance.
For the two baseline results, a small but noticeable increase in accuracy is observable in case of Hebbian weight selection.
It is noteworthy, that -- despite this rather minor influence on the overall classification accuracy -- \gls{stdp} constitutes the driving force for the expression of meaningful synapses in the presented implementation of structural plasticity.

\begin{figure}[t]
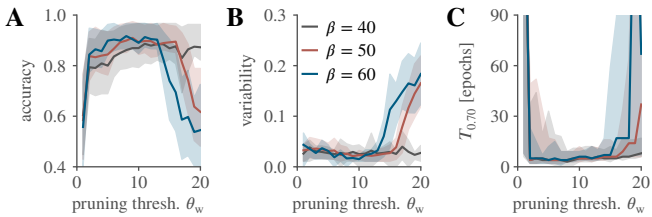

	\begin{center}
		\scalebox{\FigureScale}{
			\begin{tikzpicture}
				\node[anchor=north west] at (0,0) {\input{figures/threshold_sweep_accuracy_paper.pgf}};
				\node[anchor=north west] at (2.9,0) {\input{figures/threshold_sweep_variability_paper.pgf}};
				\node[anchor=north west] at (5.8,0) {\input{figures/threshold_sweep_convergence_paper.pgf}};
				\node at ($(0.0,0) + (0.11,-0.30)$) {\small\bfseries A};
				\node at ($(2.9,0) + (0.11,-0.30)$) {\small\bfseries B};
				\node at ($(5.8,0) + (0.11,-0.30)$) {\small\bfseries C};
			\end{tikzpicture}
		}
	\end{center}
	\caption{
		\tb{Stability of network performance over a wide range of hyperparameters.}
		We varied the pruning threshold $\theta_\text{w}$ and the regularization strength $\beta$, which both shape the steady-state weight distribution.
		For different aspects of learning performance, broad plateaus with respect to variations of these hyperparameter can be observed.
            	Solid lines and shaded areas respectively denote mean and 20-80 percentiles, measured over 20 randomly initialized experiments.
		The plateaus mostly coincide for
		\tb{(A)} classification accuracy after learning (average over the last 20 epochs),
		\tb{(B)} variability of accuracy after learning (standard deviation over the last 20 epochs), and
		\tb{(C)} number of epochs until an accuracy of \SI{70}{\percent} was reached.
		}
	\label{fig:hyperparameter}
\end{figure}

The network's performance depends on the selection of hyperparameters for the learning rule.
Since the pruning condition is based on the synaptic weights, the selection of the pruning threshold must take into account the distribution of learnt efficacies (Fig.~\ref{fig:weight_distribution}).
Thus, $\thetaw$ must be high enough to allow uninformative synapses to be pruned, but still low enough as to not affect previously found informative synapses.
Fig.~\ref{fig:hyperparameter} displays different performance metrics as a function of the pruning threshold.
These analyses are shown for a varied strength of the regularizing term $\beta$, as the weight distribution and scale depend on the balance of the positive Hebbian and this negative force.
All three metrics exhibit broad plateaus of good performance, which coincide over a relatively wide range of $\thetaw$.

\subsection{On-the-fly adaptation to switching tasks}

As demonstrated, structural plasticity enables learning in sparse networks by exploring the input space and forming informative receptive fields.
So far we have considered experiments with a randomly initialized connectome and most importantly a homogeneous weight distribution.
In another experiment, we tested the plasticity mechanism's ability to cope with a previously learned and therefore already structured weight distribution.
We achieved this by abruptly changing the task during training.
After 200 epochs, the receptors were moved to new, random locations, resulting in a misalignment of receptive fields and data points.
The plasticity rule was executed continuously, before and after this task switch.

\begin{figure}[t]
	\begin{center}
		\scalebox{\FigureScale}{
			\input{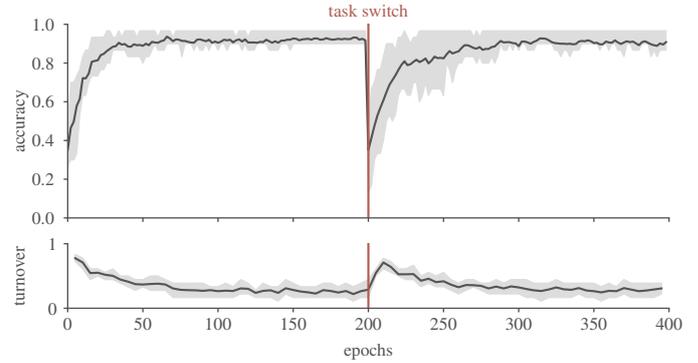}
		}
	\end{center}
	\caption{
		\tb{Restoration of network performance after task switch.}
		After training for 200 epochs, the receptor layer is randomly rearranged, leading to a mismatch in receptive fields.
		Ongoing structural plasticity unlearns the previously established connectome and quickly starts to again explore the input space.
		This process can be observed in an elevated turnover rate after the task switch, similar to the initial phase of the experiment.
		}
	\label{fig:lesion}
\end{figure}

As shown in Fig.~\ref{fig:lesion}, the accuracy dropped to approximately chance level as the receptors were shuffled.
This decline, however, was directly followed by a rapid increase of the turnover rate.
The negative contribution of the regularization term outweighed the Hebbian forces, thereby resulting in decreasing synaptic efficacies.
After a few epochs, most of the weights had fallen below $\theta_w$ and were eligible for pruning.
This process allowed the network to successfully unlearn previous connections, thus rekindling exploration of the input space.

\subsection{Fast and efficient hardware emulation}

In our proposed implementation, structural reconfiguration only induces a small computational overhead.
Synaptic pruning and reassignment is enabled by exploiting the synaptic filtering of spike events by their source address.
Since the connectome is essentially defined by the address labels stored in the synapses' memory, it can also be reconfigured with local operations only.

The algorithm executed on the on-chip microprocessor can effectively be dissected into four steps (Alg.~\ref{alg:plasticity_algorithm}):
accessing the synaptic weights, evaluation of the pruning condition, potential reassignment of the synaptic label, and a final write access to the synapse \gls{sram}.
The exact time required for executing the respective instructions depends on the neuromorphic system's architecture and the design of the plasticity processing unit.
In general, memory access and the generation of pseudo-random numbers can be regarded as the most expensive operations.
The former primarily depends on the system's design and can be optimized for low access times.
Random number generation can also be sped up by implementing dedicated hardware accelerators.

\begin{figure}
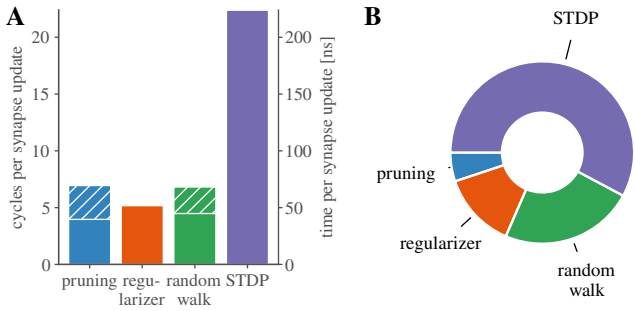

	\begin{center}
		\scalebox{\FigureScale}{
			\begin{tikzpicture}
				\node[anchor=north west] (a) at (0.0, 0.0) {\input{figures/performance/timing_bar.pgf}};
				\node[anchor=north west] (b) at (4.7, 0.0) {\input{figures/performance/timing_pie.pgf}};

				\node at ($(a.north west) + (0.22,-0.30)$) {\small\bfseries A};
				\node at ($(b.north west) + (0.22,-0.30)$) {\small\bfseries B};
			\end{tikzpicture}
		}
	\end{center}
	\caption{
		\tb{Efficient mixed-signal implementation of structural plasticity.}
		\tb{(A)} Duration of a synapse update broken down into its four individual contributions, including structural reconfiguration.
		The hatched areas indicate the time spent on pseudo-random number generation.
		\tb{(B)} Contributions of the individual terms to the overall update duration, taking into consideration that pruning and reassignment are executed five times less often than synaptic weight updates.
		}
	\label{fig:timing}
\end{figure}

Our implementation on BrainScaleS-2 is enabled by the \gls{ppu} and its tight coupling to the neuromorphic core.
Access to the synapse array as well as arithmetic operations are optimized by a parallel processing scheme.
Performing a structural plasticity update on a single slice of 16 synapses takes approximately 110 clock cycles, which corresponds to \SI{1.1}{\micro\second} at a \gls{ppu} clock frequency of \SI{100}{\mega\hertz} (Fig.~\ref{fig:timing}).
This amounts to about seven clock cycles, or \SI{69}{\nano\second}, per synapse.
In comparison, the Hebbian term, which is executed five times more often, requires approximately \SI{3.8}{\micro\second} for a slice or \SI{240}{\nano\second} per synapse.
The regularizer and random walk take \SI{69}{\nano\second} and \SI{97}{\nano\second} per synapse, respectively.
In our implementation, these terms were implemented separately and were not particularly optimized for performance.
Sharing memory accesses or intermediate results between them would lead to an overall speedup of the plasticity mechanism.

The time spent on the generation of pseudo-random numbers, highlighted in Fig.~\ref{fig:timing}, constitutes a significant portion for both the random walk and the pruning term.
On the full-size BrainScaleS-2 system, hardware accelerators allow to reduce this contribution to a comparatively negligible 0.08 clock cycles per synapse\footnotemark[2].

Hence, our implementation of structural plasticity is doubly efficient.
Not only can it effectively optimize the utilization of synaptic resources, but it can also achieve this at the cost of only a small overhead to the calculation of synaptic weight updates (Fig.~\ref{fig:timing}).

The accelerated nature of the BrainScaleS-2 system also contributes to a rapid evaluation of plasticity schemes in general -- and structural reconfiguration in particular.
Emulating a single epoch of 24 biological seconds required a total of \SI{137}{\milli\second} on our system.
Excluding the overhead induced by on-the-fly generation of input spike trains in Python, this number boils down to less than \SI{50}{\milli\second}, which corresponds to a speedup factor of about 500.
As shown by \citet{wunderlich2019demonstrating}, this overhead can be dramatically reduced by porting the experiment control from the host and \gls{fpga} to the \gls{ppu}.
This further allows to optimize the system's power consumption to below \SI{60}{\milli\watt}, with only a weak dependence on the nature of ongoing network activity and plasticity \citep{wunderlich2019demonstrating}.

BrainScaleS-2 achieves its speedup by exploiting the quick emulation with above-threshold analog transistor circuits.
The individual parametrization of each circuit allows to dramatically reduce fabrication-induced fixed-pattern variations~\citep{aamir2018lifarray}.
We employed calibration routines to equilibrate the behavior of the synaptic correlation sensors as well as the neuron circuits.
To assess the remaining variations, we investigated the transferability of the network's topology and weight information.
For this purpose, we trained a population of three label neurons and then replicated the learnt connectome to four other groups of neurons and their associated synapse columns.
Since each of these instances of the network exhibited its own intrinsic circuit variations, this allowed us to infer a measure of transferability of training result across systems.
The accuracies acquired for these networks deviated by only \SI{1.2}{\percent} from the originally trained population.
This shows that a calibrated system can be used for inference with weights learnt on a different setup.
Nevertheless, learning can partially compensate for non-ideal calibration data~\citep{wunderlich2019demonstrating}, stressing the value of on-chip training.

\section{Discussion}
We have presented a fully local, on-chip structural plasticity mechanism together with an efficient implementation on a prototype of the BrainScaleS-2 architecture.
The algorithm allows to train a network with a sparse connectome, thereby utilizing synaptic resources more efficiently.
We showcased this implementation in a supervised learning task with weight updates driven by Hebbian potentiation.
For this classification task, it was possible to drastically increase the sparsity of the connectome without significant performance loss.
Self-configuring receptive fields led to near-perfect accuracy and a better utilization of synaptic resources without prior knowledge of the input data.

Structural plasticity has already been successfully applied to networks with various topologies and learning paradigms \citep{butz2009activity,george2017,bogdan2018structural,kappel2015synaptic,bellec2017deep}.
In addition to a more efficient handling of synaptic resources, it is assumed to also improve network performance~\citep{roy2014liquid,spiess2016structural} and, in conjunction with non-linear multi-compartmental neuron models, memory capacity~\citep{poirazi2001impact,hussain2016multiclass}.
We expect our plasticity scheme to be applicable to many of these different network topologies and learning rules:
The weight dynamics can be easily extended by additional terms, e.g. modulatory reward signals, as they were used by \citet{wunderlich2019demonstrating}.
Alternatively, the proposed implementation of pruning and reassignment could be combined with completely different weight update mechanisms.
This would allow to alleviate the ubiquitous issue of limited fan-in for multi-layer networks, that have, on their own, already been demonstrated on BrainScaleS~\citep{schmitt2017neuromorphic,kungl2018generative}.
Furthermore, as the \gls{ppu} is freely programmable, the structural plasticity mechanism itself can be extended by e.g. additional pruning criteria such as bookkeeping~\citep{spiess2016structural}, spatial information~\citep{bogdan2018structural}, or silent synapses~\citep{roy2016online}.
However, all of these additions should be considered regarding their impact on the algorithm's locality.

There already exist several implementations of structural plasticity for various neuromorphic platforms.
Most of them were based on an on-the-fly adaptation of connectivity tables.
Such generally very flexible strategies have been successfully demonstrated especially on digital systems~\citep{bogdan2018structural,yan2019efficient}, where the event handling per se is already centered around look-up tables.
For such an approach, the ordering of connectivity lists is important to minimize look-up latencies, which introduces overhead for the removal and insertion of a synapse~\citep{liu2018memory}.
Related strategies were proposed also for analog neuromorphic systems~\citep{spiess2016structural,george2017,bhaduri2018spiking}.
These implementations were based on optimized look-up matrices, using a representation comparable to our on-chip synapse matrix, which were stored and evaluated on external \glspl{fpga}.
In these cases, learning and rewiring were also executed off-chip.
In contrast, BrainScaleS-2 provides a local, in-synapse definition of the sparse connectome.
This allowed our efficient implementation of on-chip plasticity and rewiring.

All of the named approaches have to allocate memory besides the actual synaptic weights, as sparse matrices always require the annotation of the placement of non-zero elements.
However, the additional increase in memory is outweighed by the overall gains due to the smaller network graphs.
External look-up tables can often be stored in \gls{dram}, which reduces their spatial footprint compared to \gls{sram}-based implementations.
The inherent access latencies can, however, be detrimental, especially for accelerated neuromorphic systems.


We note that the accelerated nature of the BrainScaleS-2 system is especially relevant in the context of modeling biological rewiring processes.
In vivo, structural changes to the connectome typically take place on time scales of hours to days \citep{lamprecht2004structural}, which allows synapses to process large amounts of information and evolve accordingly before being potentially pruned.
This throughput of information -- essentially spikes -- per unit of time is directly contingent on the specific time constants of neuro-synaptic dynamics.
Consequently, the acceleration factor of BrainScaleS-2 can also translate directly to a corresponding speedup of structural plasticity.

Our implementation scales well with growing system sizes, since it is fully based on synapse-local quantities.
In particular, it profits directly from the parallel handling of synaptic updates.
On large systems, this would especially benefit the more complex network structures and associated larger synapse arrays required when tackling more difficult tasks.

\section*{Contributions}
SB and BC conceived the idea and designed, implemented, and executed the experiments.
MP contributed to the experiment design as well as the evaluation and interpretation of the results.
KS designed the peripheral hardware for the experiment setup.
JS, as the lead designer and architect of the neuromorphic platform, conceived and implemented the synapse circuits.
All authors contributed to the manuscript.

\section*{Acknowledgements}
The authors express their gratitude towards A. Baumbach, E. Müller, P. Spilger, and Y. Stradmann for their support and helpful discussions.\\
\newline
\noindent
This work has received funding from the European Union Sixth Framework Programme (FP6/2002-2006) under grant agreement no. 15879 (FACETS),
the European Union Seventh Framework Programme (FP7/2007-2013) under grant agreement no. 604102 (HBP), 269921 (BrainScaleS) and 243914 (Brain-i-Nets)
and the Horizon 2020 Framework Programme (H2020/2014-2020) under grant agreement no. 720270 and 785907 (HBP), as well as the Manfred St\"{a}rk Foundation.

	\bibliography{bibliography}

\end{document}